\documentclass[10pt,twocolumn]{IEEEtran}
\IEEEoverridecommandlockouts
\usepackage{epsfig}
\usepackage{amsmath}
\usepackage{theorem}
\usepackage{amsmath} 
\usepackage{mathrsfs}
\usepackage{amsmath}
\usepackage{amsfonts}
\usepackage{amssymb}
\usepackage{mathrsfs} 
\usepackage{euscript}
\usepackage{graphicx}
\usepackage{multirow}
\usepackage{algorithm}
\usepackage{rotating}
\usepackage{subfigure}
\usepackage{cite}
\usepackage{url}
\usepackage{enumitem}
\usepackage{color}
\usepackage{float}

\usepackage{amsmath}
\usepackage{algorithm} 
\usepackage{algorithmic}   
\usepackage[algo2e]{algorithm2e} 
\begin{document}
	
		\title{\huge Toward a Smart Resource Allocation Policy via Artificial Intelligence in 6G Networks: Centralized or Decentralized?  
		\author{ Ali Nouruzi, Atefeh Rezaei, \textit{Graduate Student Member, IEEE}, Ata Khalili, \textit{Member, IEEE},
			Nader Mokari, \textit{Senior Member, IEEE},
			Mohammad Reza Javan, \textit{Senior Member, IEEE}, Eduard A. Jorswieck \textit{Fellow, IEEE}, and \\Halim Yanikomeroglu, \textit{Fellow, IEEE}
	}}
	\maketitle

\begin{abstract}
In this paper, we design a new smart software-defined radio access network (RAN) architecture with important properties like flexibility and traffic awareness for sixth generation (6G) wireless networks. 
In particular, we consider a hierarchical resource allocation framework for the proposed smart soft-RAN model, where the software-defined network (SDN) controller is the first and foremost layer of the framework. This unit dynamically monitors the network to select a network operation type on the basis of distributed or centralized resource allocation architectures to perform decision-making intelligently.
In this paper, our aim is to make the network more scalable and more flexible in terms of achievable data rate, overhead, and complexity indicators. To this end, we introduce a new metric, throughput overhead complexity (TOC), for the proposed machine learning-based algorithm, which makes a trade-off between these performance indicators. 
In particular, the decision making based on TOC is solved via  deep reinforcement learning (DRL), which determines an appropriate resource allocation policy.
 Furthermore, for the selected algorithm, we employ the soft actor-critic method, which is more accurate, scalable, and robust than  other learning methods.~Simulation results demonstrate that the proposed smart network achieves better performance in terms of TOC  compared to fixed centralized or distributed resource management schemes that lack dynamism. Moreover, our proposed algorithm outperforms  conventional learning methods employed in other state-of-the-art network designs.
	\end{abstract}


\section{Introduction }
%\subsection{Background}
The evolution from fifth-generation (5G) to sixth-generation (6G) wireless networks has proceeded along two main lines, which consist of evolving network architecture and evolving communications technologies. As 5G and beyond networks continue to evolve, it will be possible for network architecture and communication technologies to be dynamically adapted in relation to the changing needs of the network. In such a flexible architecture, a huge amount of signaling and computational resources are needed to manage the network resources efficiently. 

Due to 
discrepancies between the mathematical tractability
and the exponentially greater complexity of
wireless networking, conventional convex optimization approaches are unfortunately ineffective and may not be able to fulfill 
the precise quality of service (QoS) requirements
of emerging applications.
To tackle this  issue, artificial intelligence has been identified as a promising solution for automatic and autonomous network
management. Adopting intelligent resource allocation
for wireless networks not only has the potential
to replace the manual intermediation needed for current network management
tasks, but also presents novel 
optimization possibilities to ameliorate performance gains in real-time.~Since future dense wireless networks will involve high complexity algorithms, machine learning (ML) has emerged as a key enabler to manage high complexity for real-time
implementation. In this regard, reinforcement learning (RL) as a type of ML has been employed to learn from an environment by trial and error, which promotes improvements over time. Also,  deep reinforcement learning (DRL)  has been investigated for comprehensive inputs as well as more accurate results in comparison to  RL algorithms \cite{mag}. Furthermore, to enable a real-time scheduler for stochastic environments, it has been shown that learning via multiple agents can solve complicated stochastic optimization problems more practically\cite{SAC_mec}.~Although DRL methods have been applied to several resource allocation problems, they have two major challenges: high sample complexity and sensitivity to hyper-parameters\cite{SAC}.  
To address these issues, the soft actor-critic algorithm with a maximum entropy objective can
be leveraged to provide more accurate and stable solutions for  scalable networks in dynamic 
environments.
 
With the development of 6G, the baseband function placement
in cloud radio access network (C-RAN) can be performed by the software-defined networking (SDN) control/management
plane based on a complete knowledge of the network. Also, 
in the open RAN architecture, this controller could be implemented as a RAN intelligent controller (RIC), which is responsible for controlling and optimizing RAN functions.
However, since the aggregation nodes in the access network evolve
over time, the knowledge of the network state requires
frequent interactions between the controller and network's entities. It may be desirable to have latency insensitive tasks performed by a centralized unit, while
relatively more stringent latency constraints
can be performed at edge nodes.
Additionally, centralized approaches are not scalable enough to
satisfy computational requirements of  networks with huge dimensions. To address this, distributed algorithms can be adopted that provide more scalable solutions when adding or removing virtual or
physical baseband resources.  The high level of flexibility
envisioned for 6G can be exploited by designing algorithms that
activate baseband functions on demand. As a result of the architectural evolution and flexibility requirements, future  network should be able to dynamically adopt the proper resource management strategy, whether it be centralized or distributed. In addition, the choice of transmission technology adopted could take into account the dynamics of the environment as well as user density and their
traffic volume. Consequently,
to guarantee real-time services, it is more beneficial to consider both centralized and distributed frameworks depending on network status and QoS requirements. To illustrate this, 
this paper proposes a new learning based resource allocation framework that considers network status, resource budget, and loads to determine the allocation policy.~In fact, in the literature, %the resource management is performed based on the centralized or decentralized in a default manner where no  software-defined networking decision is investigated to determine the network's operation mode.
resource management is typically either centralized or decentralized by default, and software-defined networking decisions have not been investigated to determine the best operation mode
% The aim of this paper is therefore to raise and discuss the question: What if 6G would be designed in a smart manner that could dynamically and intelligently switch between centralized and distributed network operations?
Our research therefore considers the implications for performance of a 6G network that can dynamically and intelligently switch between centralized and distributed network operations.

%~Being motivated by the above issue, in this paper, we aim to provide an smart cooperation between the centralized controller and SDN based on the smart resource allocation policy.

%Multi agent double deep reinforcement learning (MA-DDRL) is one of machine learning algorithm in that each agent or nodes by interacting with environment train themselves how to reach to best policy. %Generally, the MA-DDRL contains two deep Q-network (DQN) and based on the architecture of the network, one of them can be located in the center node and another one can be deployed distributively. 
%Federated learning (FL) is a kind of machine learning which enable learning in a distributed way without exchanging all of the data between the center node and local node. Hence, by preserving privacy, reduce the usage of backhaul and latency.
\subsection{Resource Allocation for Wireless Communication}
Let us start by introducing the candidate transmission strategy and the resource allocation frameworks in  conventional network architectures.~The idea of improving the performance of RAN architecture in 5G and beyond has been considered in the literature, where power domain non-orthogonal multiple access (PD-NOMA) is one of the best candidates for for resource allocation policies. Furthermore, several works have tried to model the structure and efficiency of the network the structure and efficiency of the network by centralizing functionality in a baseband unit (BBU) pool \cite{IAB,Ahmad}. In \cite{multiagent}, a distributed structure was proposed to improve the latency and proficiency of a network.

%Moreover, there are some works that combine the centralized case with  a {distributed} one  \cite{zhang2019deep,Xia1,atefeh,jointcendes}.
%\footnote{It is worth mentioning we classified these papers in to semi-centralized scenario based on our assumption about ideas of centralized, semi-centralized, and distributed cases. In our paper, the semi-centralized scenario happen when resource allocation tasks are performed by BSs with some signaling with BBU pool and distributed scenario is the case when the whole resource allocation tasks are performed by BSs without any singling with other nodes.  The authors of these papers, consider the distributed scenario against the centralized case, where part or all tasks of resource allocation are performed by a part, other than centralized processor with some singling.}.
% Besides, AI is also considered in RAN \cite{SAC_mec,multiagent}. Based on our paper topic, we categorize the related works as follows:
%\begin{itemize}
	 
	\subsubsection{ Centralized Resource Allocation}
	In such architectures, BS information is supposed to be gathered in the centralized controller to perform the overall resource management. In fact, in this scheme, the baseband signals of
	the distributed units are processed by the central processor for the purpose of effective interference management.
	% In \cite{sepehr}, 	the authors propose an algorithm for joint power allocation, CoMP scheduling, and NOMA clustering  based on mixed integer monotonic optimization and sequential programming,	respectively.
	 In \cite{Ahmad}, a  multi-agent DRL based algorithm was employed for sum-rate maximization in a centralized network.~In \cite{Zeng2}, the authors proposed a DRL based resource allocation policy to enhance the network performance of the multi-carrier NOMA system.~However, these works employed a centralized architecture,
 which entails more signaling between BSs and the  centralized controller.\\
	%	Also, the authors in \cite{yin2020semi} considered a network heterogeneous network with a macro base station (MBS) that cooperates with  small base stations (SBS) performs a cooperative framework for semi-centralized resource allocation aiming to rate maximization.
	%The work in \cite{ekram} presents a joint resource allocation and admission control in a C-RAN.
	%In \cite{c1}, a low-complexity heuristic packet scheduling	scheme is proposed for a downlink centralized network based on the ultra-reliable low latency communications.
%	~In \cite{Hao1}, a multiple-input multiple-output (MIMO) in CoMP enabled-remote radio head cluster (RRHC) is considered  to maximize the downlink weighted sum rate. %In \cite{Elhattab1}, the authors consider the heterogeneous cloud radio access network (HC-RAN) to maximize the throughput of the network by considering device association, radio resource allocation, and power allocation.  
%	Also, the authors in \cite{Moltafet1} investigate a C-RAN based on a sparse code multiple access (SCMA) scheme. They propose a robust method for the joint codebook allocation and user association subproblem.

		\subsubsection{Distributed Resource Allocation}
		In this scheme, resource allocation and management are performed in a distributed manner on the basis of information available locally.
			    % The authors in \cite{wang2016} consider a joint resource scheduling scheme in C-RAN regarding both the computation resources in BBUs and the antenna resources provided by RRHs.
		%~Moreover, the joint congestion 	control and resource allocation to balance tradeoff between throughput utility and delay performance is studied in \cite{li23}.
		% In \cite{}, the resource allocation for the Internet of Thing (IoT) applications with latency requirements in fog-RAN is investigated. The authors employ the RL method to solve the Markov decision process (MDP) problem. % In \cite{Hua1}, the network slicing in 5G communication systems is studied in which the resource allocation problem is solved via DRL based algorithm.
		  The authors in \cite{multiagent} developed a multi-agent
	DRL based resource allocation policy 
	for the heterogeneous  QoS requirements in vehicular networks. In \cite{Guo}, the authors proposed an iterative algorithm for subcarrier
	assignment and power allocation by using a DRL based method that considered the impact of SIC errors.~However, in this scheme, BSs performed local signal processing which degrades the performance of the dense networks due to incorporating interference. \\

	\subsubsection{Combination of Centralized and Distributed Resource Allocation}
	
	 In \cite{SAC_mec}, a multi-access edge computing technique was considered that reduced the core network congestion. More specifically, cooperative computation offloading policy was designed for MEC technology using the soft actor-critic (SAC) method for both the centralized and distributed offloading.
	In \cite{zhang2019deep}, a DRL algorithm for vehicular-to-everything (V2X) communication was proposed that determined resource block allocation and performed power control. The DRL algorithm also selected the transmission mode (i.e., vehicle-to-infrastructure [V2I] or vehicle-to-vehicle [V2V] communications.
	 The authors in \cite{Xia1} performed functional splits
  of control and data planes between the cloud and edge nodes in C-RAN while taking the fronthaul delay into account.
% In \cite{Kang1}, the cloud-edge allocation of the control function of rate selection and the data plane function of decoding on uplink communication in the Fog-aided network architectures is proposed.
 The authors in \cite{atefeh} considered a semi-centralized framework for the resource allocation problem by using matching theory and a successive convex approximation (SCA) approach.
Similarly, the authors in \cite{IAB}  proposed a semi-centralized resource allocation scheme to maximize the weighted matching (MWM) problem  for  integrated access and backhaul (IAB) networks.
		The authors in \cite{jointcendes} compared both centralized and distributed algorithms regarding the BBU hotel location problem in C-RAN where their proposed solution is based on a distributed heuristic algorithm.
		Moreover, in \cite{valaee1}, the authors compared the energy efficiency of their proposed distributed and centralized user association algorithms by sequentially minimizing the power consumption of the heterogeneous network they considered.
	%	  all of the above-mentioned works assume that the resource allocation is only performed in either centralized or distributed architectures without considering any opportunity to change the solution structure. 
	In each of the aforementioned works, resource allocation was performed in either a centralized or distributed framework; none of the works considered the possibility of dynamically changing the framework.
%	\end{itemize}
	A summary of these works is shown in Table \ref{Tab:REF}.

	\begin{table*}[t]
	\caption{Summary of related works and the contribution of this paper in relation to them.}
	\label{Tab:REF}
	\tiny  
	\begin{center}
		\resizebox{18cm}{!} 
		{ 
			\begin{tabular}{ | p{1cm} |  p{12cm} | }
				\hline
				\textbf{Ref.} &  \textbf{Solution method}  \\ \hline
				%       	 & communication, computing resource allocation, user association, and BBU and BS mapping & Delay minimization of small cell users &  Auction theory  & Distributed \\ \hline
				%		 \cite{Zhang4}& Task offloading &Energy efficiency& Lyapunov optimization theory, convex
				%		 decomposition methods and matching game& Semi-centralized\\ \hline
				%	\cite{c1}& a heuristic packet scheduling algorithm for centralized network (conventional solution)\\
				%	\hline
				\cite{SAC_mec} &   Offloading at the MEC via an soft actor-critic algorithm (AI solution)\\ \hline
				\cite{IAB} &  Maximum weighted matching for semi-centralized resource management (conventional solution) \\ \hline 
				% \cite{sepehr} & A monotonic optimization and sequential programming  algorithm for multi-cell NOMA network (conventional solution)\\
				%	\hline
				\cite{multiagent}&   A multi-agent
				deep reinforcement learning (MADRL) method for networks with heterogeneous services (AI solution)\\ \hline 
				\cite{Ahmad} & A DRL algorithm for multi-cell and multi-user (AI solution)\\
				\hline
				%		\cite{} &  \\ \hline
				% \cite{li23} & Lyapunov optimization technique and the Lagrange dual decomposition method in a distributed network (conventional solution)\\ \hline
				\cite{Zeng2} &  DRL for multi-carrier-single-cell NOMA system (AI solution) \\ \hline
				\cite{Guo} &  Iterative resource allocation algorithm based on DRL (AI solution) \\ \hline
				%		\cite{Zhang3}&Sub-carrier assignment&Maximize the system capacity&Matching theory and coalition game theory& Distributed \\ \hline
				\cite{zhang2019deep} & DRL in a decentralized structure based on cellular V2X	communications (AI solution) \\ \hline
				\cite{Xia1} &   Matroid
				constrained sub-modular maximization problem and heuristic algorithms in  a centralized-distributed network (conventional solution)\\ \hline
				\cite{atefeh} &   Matching theory and SCA in semi-centralized network (conventional solution)\\ \hline
				\cite{jointcendes} & Comparison performance of distributed and centralized frameworks via distributed heuristic algorithm (conventional solution) \\ \hline 
				Our paper & AI method based on smart solution centralized/distributed (AI solution and soft actor-critic based method) \\ \hline 
			\end{tabular}
		}
	\end{center}
\end{table*}
\begin{table*}[]
	\centering
	\begin{tabular}{|c|ccc|cccc|}
		\hline
		\multirow{2}{*}{\textbf{Ref.}} & \multicolumn{3}{c|}{\textbf{AI Methods}}                            & \multicolumn{4}{c|}{\textbf{Policy on RA}}                                            \\ \cline{2-8} 
		& \multicolumn{1}{c|}{SAC} & \multicolumn{1}{c|}{DDPG} &  \multicolumn{1}{c|}{MARL} & \multicolumn{1}{c|}{{Distributed }} & \multicolumn{1}{c|}{Centralized} & \multicolumn{1}{c|}{{Semi-Centralized }} & \multicolumn{1}{c|}{Smart} \\ \hline
		\cite{SAC_mec}		& \multicolumn{1}{c|}{$\color{green}{\checkmark}$} & \multicolumn{1}{c|}{$\color{red}{\pmb{\mathsf{x}}}$} & $\color{red}{\pmb{\mathsf{x}}}$ & \multicolumn{1}{c|}{$\color{green}{\checkmark}$} & \multicolumn{1}{c|}{$\color{green}{\checkmark}$} & \multicolumn{1}{c|}{$\color{red}{\pmb{\mathsf{x}}}$} & $\color{red}{\pmb{\mathsf{x}}}$ \\ \hline
		\cite{Ahmad}& \multicolumn{1}{c|}{$\color{red}{\pmb{\mathsf{x}}}$} & \multicolumn{1}{c|}{$\color{red}{\pmb{\mathsf{x}}}$} & $\color{green}{\checkmark}$ & \multicolumn{1}{c|}{$\color{green}{\checkmark}$} & \multicolumn{1}{c|}{$\color{green}{\checkmark}$} & \multicolumn{1}{c|}{$\color{red}{\pmb{\mathsf{x}}}$} & $\color{red}{\pmb{\mathsf{x}}}$ \\ \hline
		\cite{Zeng2}	& \multicolumn{1}{c|}{$\color{red}{\pmb{\mathsf{x}}}$} & \multicolumn{1}{c|}{$\color{red}{\pmb{\mathsf{x}}}$} & $\color{red}{\pmb{\mathsf{x}}}$ & \multicolumn{1}{c|}{$\color{red}{\pmb{\mathsf{x}}}$} & \multicolumn{1}{c|}{$\color{green}{\checkmark}$} & \multicolumn{1}{c|}{$\color{red}{\pmb{\mathsf{x}}}$} &  $\color{red}{\pmb{\mathsf{x}}}$\\ \hline
		
		\cite{multiagent}  & \multicolumn{1}{c|}{$\color{red}{\pmb{\mathsf{x}}}$} & \multicolumn{1}{c|}{$\color{red}{\pmb{\mathsf{x}}}$} & $\color{green}{\checkmark}$ & \multicolumn{1}{c|}{$\color{red}{\pmb{\mathsf{x}}}$} & \multicolumn{1}{c|}{$\color{green}{\checkmark}$} & \multicolumn{1}{c|}{$\color{red}{\pmb{\mathsf{x}}}$} & $\color{red}{\pmb{\mathsf{x}}}$ \\ \hline
		
		\cite{zhang2019deep}     & \multicolumn{1}{c|}{$\color{red}{\pmb{\mathsf{x}}}$} & \multicolumn{1}{c|}{$\color{red}{\pmb{\mathsf{x}}}$} &$\color{red}{\pmb{\mathsf{x}}}$ & \multicolumn{1}{c|}{$\color{red}{\pmb{\mathsf{x}}}$} & \multicolumn{1}{c|}{$\color{green}{\checkmark}$} & \multicolumn{1}{c|}{$\color{green}{\checkmark}$} & $\color{red}{\pmb{\mathsf{x}}}$  \\ \hline
		
		\cite{Guo} & \multicolumn{1}{c|}{ $\color{red}{\pmb{\mathsf{x}}}$} & \multicolumn{1}{c|}{$\color{green}{\checkmark}$} & \multicolumn{1}{c|}{ $\color{red}{\pmb{\mathsf{x}}}$}  & \multicolumn{1}{c|}{$\color{red}{\pmb{\mathsf{x}}}$ } & \multicolumn{1}{c|}{$\color{green}{\checkmark}$} & \multicolumn{1}{c|}{$\color{red}{\pmb{\mathsf{x}}}$ } & $\color{red}{\pmb{\mathsf{x}}}$  \\ \hline
		Our paper &\multicolumn{1}{c|}{$\color{green}{\checkmark}$} & \multicolumn{1}{c|}{$\color{green}{\checkmark}$} & \multicolumn{1}{c|}{$\color{red}{\pmb{\mathsf{x}}}$} & \multicolumn{1}{c|}{$\color{green}{\checkmark}$}& \multicolumn{1}{c|}{$\color{green}{\checkmark}$} &\multicolumn{1}{c|}{$\color{red}{\pmb{\mathsf{x}}}$} & \multicolumn{1}{c|}{$\color{green}{\checkmark}$}\\ \hline
	\end{tabular}
\end{table*}

\subsection{Contribution}
This paper proposes an intelligent approach approach for determining an effective resource allocation policy in a downlink PD-NOMA network.
% In particular, the SDN controller decides which part of the network is responsible for resource allocation based on the network conditions. 
We consider a network that can switch between  centralized and distributed operations for  resource management  on demand. 
To the best of our knowledge, a smart network architecture configuration such as this has not been studied yet.  Previous works have not considered the changing environmental conditions of networks, and thus are not appropriate for real-world scenarios. Besides, the complexity, overhead, and achievable data rate performance metrics are not considered jointly in the other existing works in the literature while it is important to consider them jointly in selecting the resource management.
The capacity of networks to learn autonomously and change their architectures to boost performance will be a critical enabler of next-generation intelligent wireless networks. The main contributions of this paper are summarized as follows:
\begin{itemize}
\item In contrast to the conventional approaches that employed  
	a framework for the resource allocation based on the analytical models, an
	intelligent approach based on the learning methods is exploited for solving the SDN decision and resource allocation problems in a centralized or distributed manner.~This approach allows a network to adapt to the environment and perform more effectively in real-world scenarios.
	\item We introduce a novel algorithm for the SDN controller. This algorithm chooses the best resource allocation policy based on the DRL method. In this framework,  centralized or distributed modes are selected on the basis of a throughput metric that considers overhead, complexity, and the total data rate. In particular, our proposed smart algorithm decides whether a BBU or each BS will be responsible for the resource allocation policy. 
	\item In the centralized scheme, we consider the single-agent SAC algorithm at the centralized unit that designates the appropriate actions based on the collected information. Additionally, in the distributed scheme, we consider that there is no information exchange between agents, and that the BSs (as RRSs) locally perform resource allocation tasks based on a multi-agent actor-critic algorithm. This algorithm can be applied to the more complex environments and it provides stable solutions for the network compared to the other learning methods.  
		\item Simulation results indicate a performance gain by employing the SAC methods relative to other ML based(i.e., DDPG and DQN approaches). Furthermore, results demonstrate performance gaps between fixed centralized or fixed distributed schemes with the proposed smart algorithm in which the time complexity, overhead, data rate are considered, simultaneously. 
		
\end{itemize}

\section{System Model and Problem Statement}
\label{sec_systemmodel}

\begin{figure*}
	\centering
	\includegraphics[width=11.00cm]{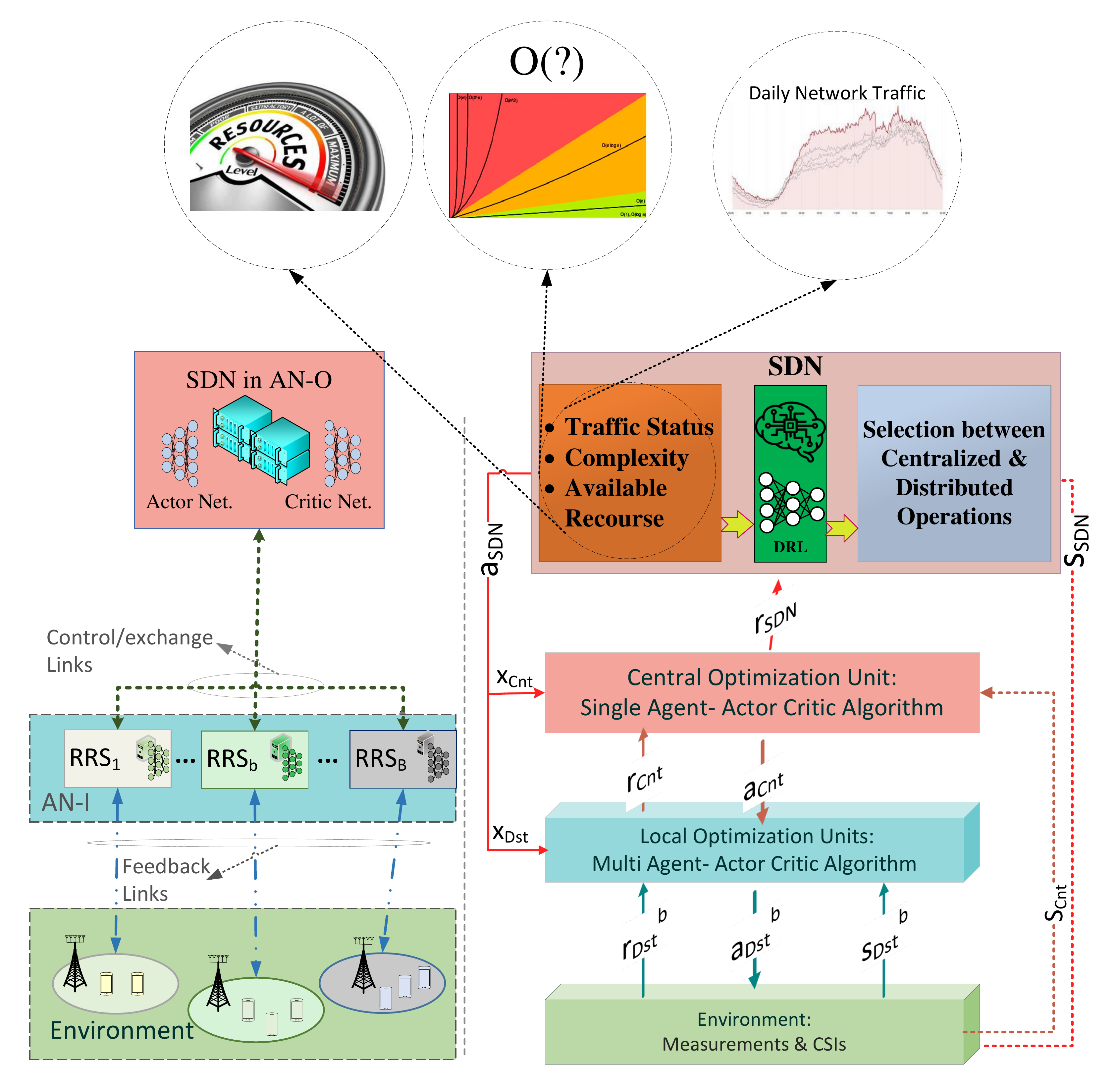}
	\caption{System model and hierarchical flowchart  of the smart soft-RAN.}
	\label{system_model}
\end{figure*}

As shown in Fig. \ref{system_model}, we consider that the RAN network consists of two major units, which we refer to as the access network outer (AN-O) and the access network inner (AN-I). The AN-O is the cloud network, and the AN-I includes $ B $ reconfigurable radio systems (RRSs) as local units. The RRSs provide signaling and data coverage for the users.  As a result of considering an intelligent network, it is necessary to obtain some efficient feedback about network status to operate the network in a real-time manner.
Furthermore, it is assumed that the information can be fed back from the BSs to the AN-O through the control links. Here we consider downlink PD-NOMA network. Also, we assume that all RRSs and users are equipped with single antennas. The AN-O consists of a BBU pool which performs centralized baseband processing, and a centralized SDN controller, which controls the network operation by programming the network's element functionalities properly.
By monitoring the network, the SDN controller determines whether the network would operate in a centralized or distributed manner.
We formulate the problem in terms of maximizing the functionality of the network. As we will see, there are two sides to this problem: decision making and resource allocation.

\subsection{Decision-Making Optimization Problem}

We consider a comprehensive framework in which the SDN controller switches between  centralized and distributed network operations by considering the total data rate and the amount of data exchanged in terms of overhead and complexity.
\\
 \subsubsection{Overhead} 
As a key performance indicator overhead is a critical key performance indicator at the considered smart network. The overhead is a function of the number of information bits needed to feed back the data of the channel status, subcarrier indicators, and the transmission power of a specific user over a subcarrier.
Also, the total number of  RRSs, users, and subcarriers in each RRS and in each time slot can affect overhead.
 
%It is considered that the fronthaul links between the BBU pool and RRSs are equipped with some optical fibers. Hence, there is always adequate bandwidth for transmitting data.
In the centralized mode, the resource allocation task is performed at the BBU, and thus the information needs to be transmitted from the RRSs to the centralized unit.  In the distributed mode, by contrast, all tasks related to  resource allocation are performed independently by the RRSs without any data exchange.\\ 
\subsubsection{Computational Complexity} 
The computational complexity of  RL-based methods depends on the number of neurons, the DNN layers, the state size, and the action space which is described in \cite{ATA}. It should be noted that the soft actor-critic method has two DNN layers and the computational
complexity of each agent as $ O_{\text{agent}} $ is a function of the complexity of both DNNs, the number of episodes, and the minibatch size. In this structure,
the complexity of each RRS is a linear function of $ O_{\text{agent}} $, the size of the subcarrier, and the users set in the RRS coverage area. \ %Moreover, the number of  output layers are set to $|A|\times|N|$ and $|A_b|\times|N_b|\times$ for the centralized method and for agent of each BS, respectively. In addition,  by increasing the number of action and state spaces number of the hidden layers of the DNN must be increases to be adopted and more applicable   \cite{xiao2019nfvdeep}.}
% \begin{figure*}[t]
% \begin{subequations}\label{problem_decision}
%\begin{align}
%\mathcal{P}_{\text{SDN}}:
%	 \max_{\boldsymbol{x}}~~   &{x}_{\text{Cnt}}^{(t)}\Big(\sum_{b\in\mathcal{B}} \sum_{n\in\mathcal{N}}\sum_{k\in\mathcal{K}} \text{log}(1+\gamma^{\text{Cnt}^{(t-1)}}_{n,b,k} )- \alpha\text{log}(\tau_{\text{Cnt}}^{(t-1)})- \zeta\text{log}(\mathcal{C}_{\text{Cnt}}^{(t-1)})\Big)	\nonumber\\ &  	+  {x}_{\text{Dst}}^{(t)}\Big(\sum_{b\in\mathcal{B}} \sum_{n\in\mathcal{N}}\sum_{k\in\mathcal{K}}  \text{log}(1+\gamma_{n,b,k}^{\text{Dst}^{(t-1)}} )- \alpha\text{log}(\tau_{b}^{(t-1)})- \zeta\text{log}(\mathcal{C}_{b}^{(t-1)})\Big),\\
%	\text{s.t.~~~}&  x_{\text{Cnt}}^{(t)}+x_{\text{Dst}}^{(t)}=1,~x_{\text{Cnt}}^{(t)},x_{\text{Dst}}^{(t)}\in\left\lbrace 0,1\right\rbrace \label{C3_decision},
%\end{align}
%\end{subequations}
%\end{figure*} 
\subsubsection{Achievable Data Rate}
  We define two integer decision making parameters for the centralized and distributed scenarios as $x_{\text{Cnt}}^{(t)}$ and $x_{\text{Dst}}^{(t)}$, respectively, where $\boldsymbol{x}^{(t)}=\{x_{\text{Cnt}}^{(t)},x_{\text{Dst}}^{(t)}\}$. To make a trade-off between data rate, overhead, and complexity, we introduce a new metric as TOC to perform mode selection by considering the data rate, overhead, and complexity functions.~Consequently, the decision-making structure for the SDN controller is shown at the top of Fig. \ref{system_model}. In this framework, we consider that in each transmission time slot $ t $ only one operation scheme can be selected. More specifically, this algorithm aims to select the best action $ a_{\text{SDN}} $ as the operation mode based on the network's traffic, where $ a_{\text{SDN}} $  determines which of $ x_{\text{Cnt}}$ and $ x_{\text{Cnt}}$ are more appropriate. It should be noted that this action is based on the determined states $ S_{\text{SDN}} $ and rewards $ r_{\text{SDN}} $ in the network. In Fig. \ref{distcent}, we define the states, actions, and rewards for the proposed resource management structure in details.

\subsection{Proposed Solution for the SDN}
\begin{figure*}
	\centering
	\includegraphics[ width=16cm]{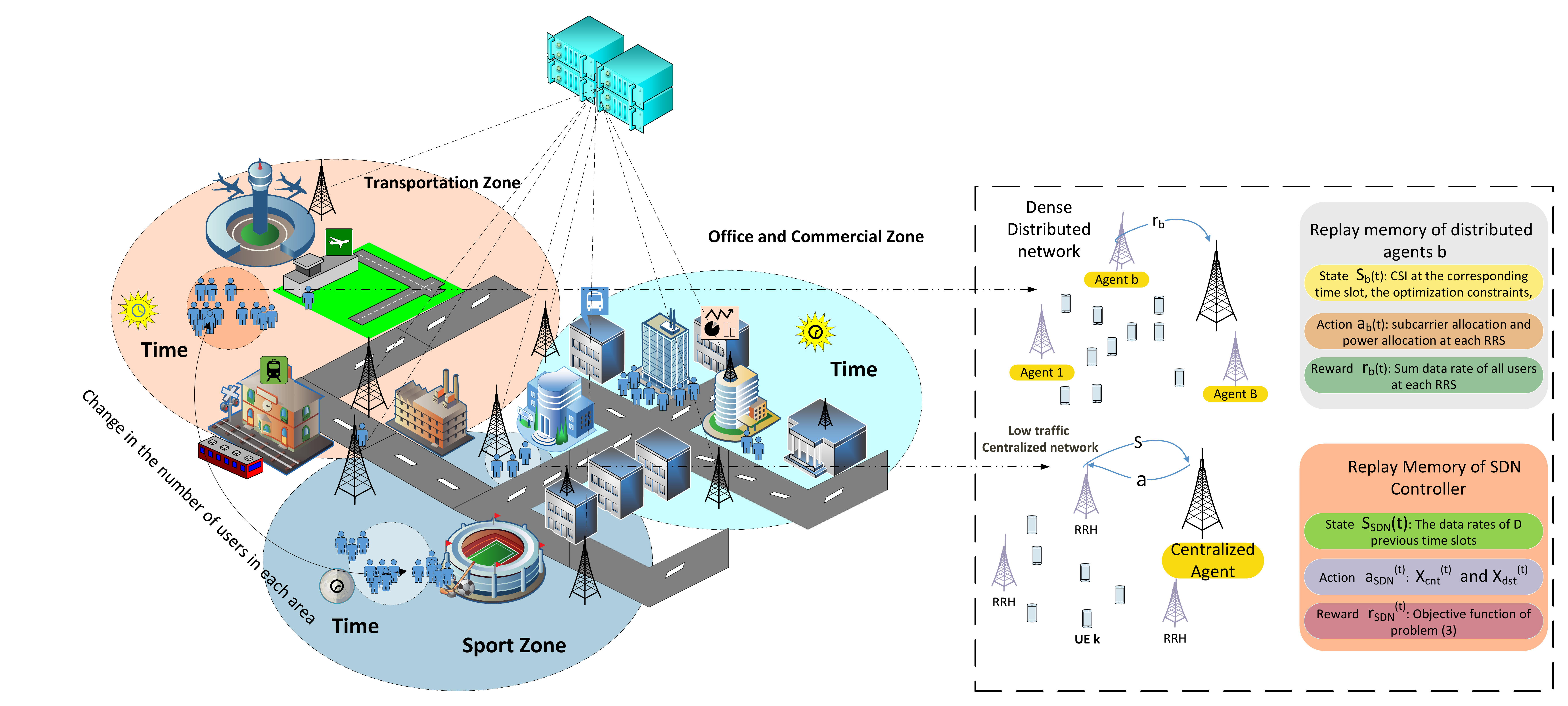}
	\caption{Network architecture based on the actor critic framework.}
	\label{distcent}
\end{figure*}

Here, we provide an efficient algorithm for deciding between the centralized and distributed schemes. The DRL framework for the decision-making problem at the SDN is employed whose states, actions, and rewards are described in Fig. \ref{distcent}. In the SDN algorithm, based on the given state $ S_{\text{SDN}}^{(t)} $ in each time slot, the SDN  selects an action $ a_{\text{SDN}}^{(t)} $. By taking the action at each time slot, the agent gets a reward $ r_{\text{SDN}}^{(t)} $  that is defined on the basis of the objective function in $ \mathcal{P}_{\text{SDN}} $. After observing $ S_{\text{SDN}}^{(t+1)} $, the SDN stores $S_{\text{SDN}}^{(t)}, a_{\text{SDN}}^{(t)}, r_{\text{SDN}}^{(t)}  $, and $ S_{\text{SDN}}^{(t+1)} $ and using Adam\cite{adam}, a method for efficient stochastic optimization, it minimizes the loss function and updates its decision.
 \subsection{Resource Allocation Problem}
After the decision-making stage, the resource allocation problem (RAP) is solved using two different schemes.~In particular, we formulate the resource allocation problem to maximize the throughput of the network while taking into account the subcarrier allocation and power control based on the downlink PD-NOMA.
To solve the problem, we propose a soft actor-critic based resource allocation for the centralized and decentralized scenarios.~In what follows, we explain how we solve the resource allocation problem.\\

\subsubsection{Centralized Scheme}   
In the case where $ x_{\text{Cnt}}=1 $, the RAP is solved on the basis on a single agent soft actor-critic at the BBU pool as shown in Fig.~\ref{system_model}. Specifically, in this scheme RRSs act as RRHs in which the radio frequency tasks and the resource allocation process are performed at the BBU pool. Each RRH collects related information and forwards it to the BBU pool. Drawing on the received information at time slot $t$, the agent chooses action $ a_{\text{Cnt}} $. Then, the actions are sent to the RRHs based on the resource allocated in the BBU pool unit. Moreover, the reward $ r_{\text{Cnt}} $ and new states $ S_{\text{Cnt}} $ are collected and forwarded to the BBU pool through fronthaul links.\\

\subsubsection{Decentralized Scheme}
\begin{figure}
	\centering
	\includegraphics[width=8cm]{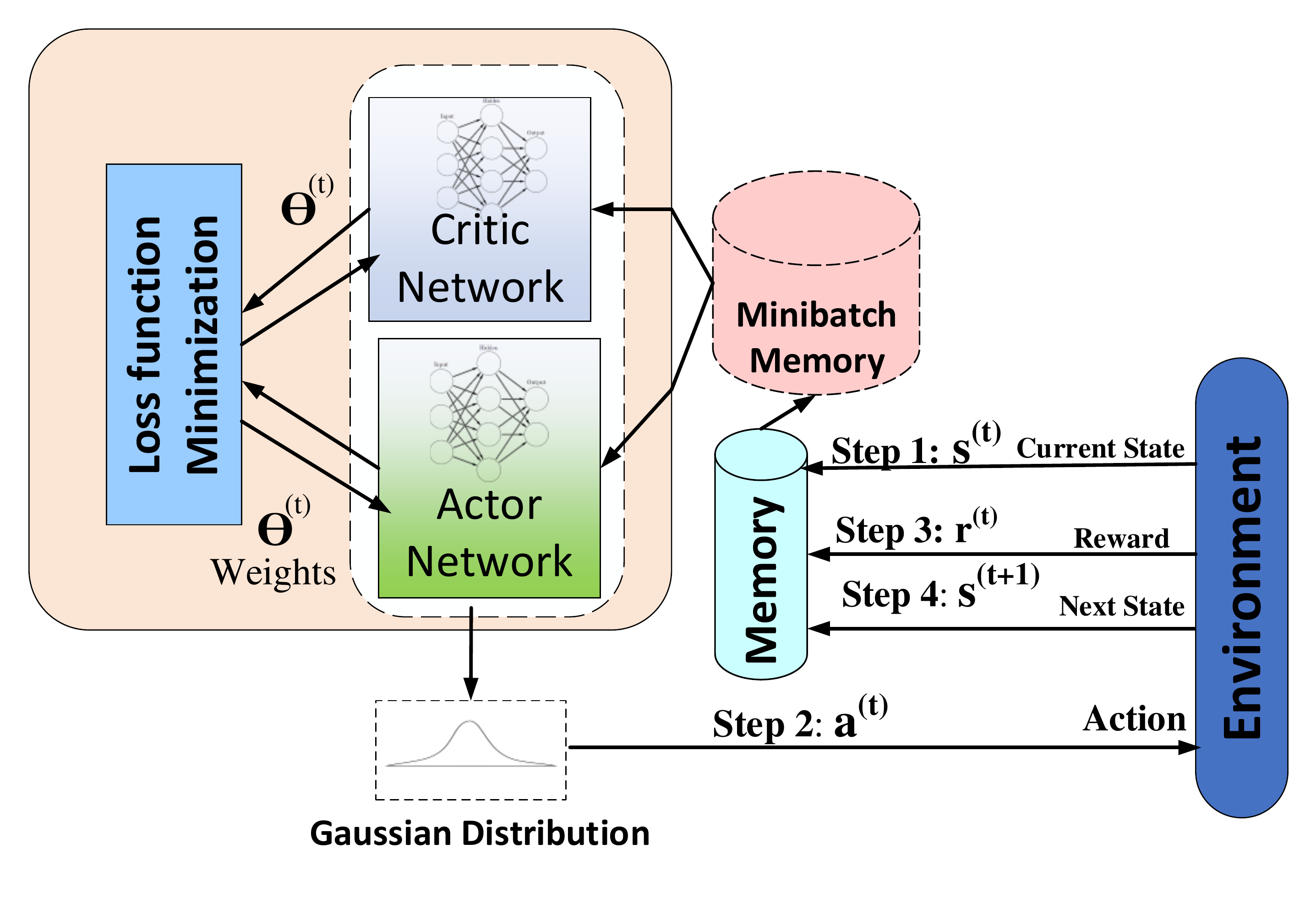}
	\caption{Actor-critic algorithm in an agent.}
	\label{DDRLF}
\end{figure}

In the decentralized scheme $\left(  \text{i.e., }  x_{\text{Dst}}=1\right)$,  resource allocation would be done by the RRSs using an independent multi-agent actor critic method.
To reduce the overhead, we consider that RRS $b$  in the distributed mode cannot access to the others' information and just performs the resource allocation tasks by using its own information.  %In this condition in addition to the state space, action space, and reward, there are two other elements, observation space $\Omega$ and observation probability $p(o^{(t)}|s^{(t)})$. Actually, the agent can only realize a partial state as an observation $o^{(t)}\in\Omega$ and by deciding on an action $a^{(t)}$, receives reward $r^{(t)}$ from the environment, and the environment moves to new state $s^{(t+1)}$ and the agent observes part of this state $o^{(t+1)}$ .  
%Belief state and belief updates are the one way to solve these kinds of problems, however there are high computational cost and memory resource \cite{Le1}. In consist the policy can be updated by using Q-learning algorithm. In this condition, the agent uses the predefined number of observations as input. By the way, some valuable observations in the past could be dismissed. To overcome this problem a recurrent neural network (RNN) is proposed \cite{Hausknecht1}. 
%Each RRS $ b $, based on the state of the environment $ S_{\text{Dst}}^b $, decides appropriate actions by using the IMA-DDRL algorithm.
 The main and target DQNs are implemented at the RRS and each agent computes the power control and subcarrier allocation locally. As we can  see in Fig. \ref{distcent}, each RRS $ b $ at time slot $t$ observes state $ S_b^{(t)}  $ and takes the action $ a_{b}^{(t)} $ individually. Also, it receives the results of its own behavior as the reward $ r_b^{(t)}  $ without knowing the actions of other agents. The proposed algorithm for each actor-critic agent is detailed in Fig. \ref{DDRLF}.

\label{Sec_Decision}

\section{Numerical Results and Discussion}
We consider a coverage area of radius $ 500$~m with four BSs distributed, each covering a radius of $100$~m.  The maximum transmit power of the BSs are set to $ 40$~dBm, and the total bandwidth is divided into $32$ orthogonal subcarriers.
The path-gain between a specific user and a BS for RF communication follows the Rayleigh distribution with path loss.  The noise power at each subcarrier is assumed $-174$ dBm/Hz.
\begin{figure*}
	\centering
	\includegraphics[width=0.7\linewidth]{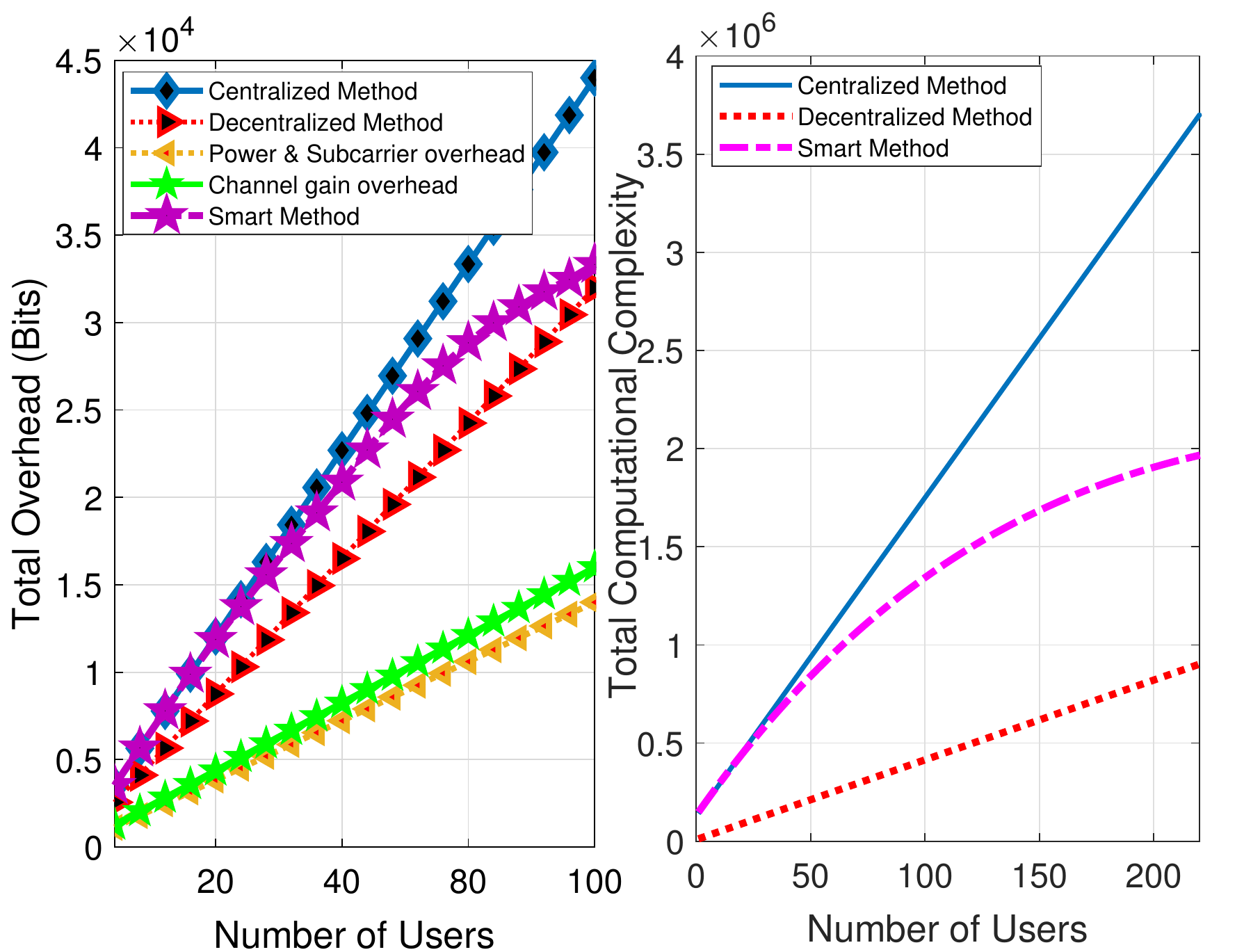}
	\caption{Total overhead and complexity relative to the number of users.}
	\label{overhead}
\end{figure*}
% Moreover, for the learning procedure, we set $ E=-- $ as the number of episodes where mini batch size is considered $ --- $.
\begin{figure*} 
	\centering
	\centerline{\includegraphics[width=10cm]{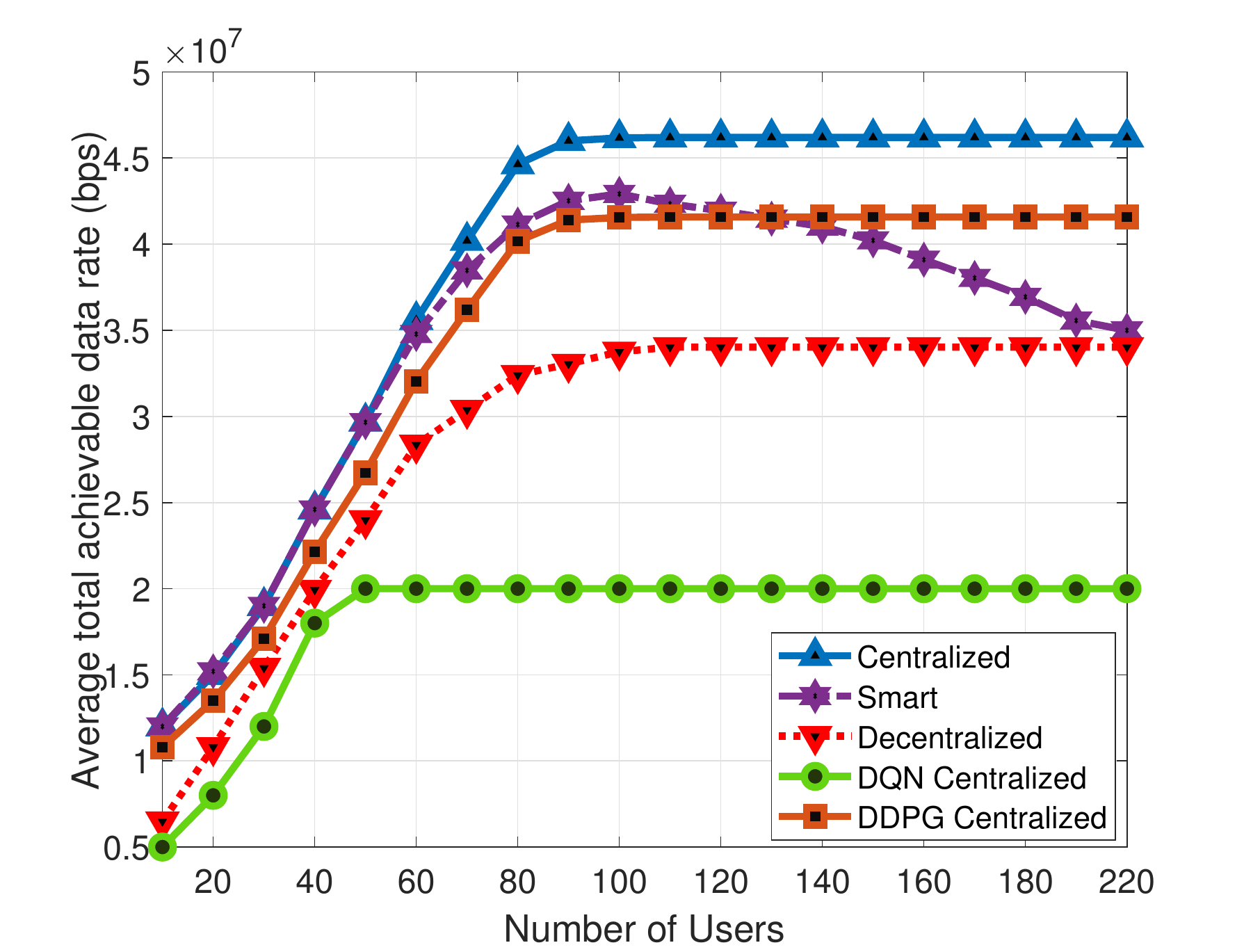}}
	\caption{Average total achievable data rate relative to the number of users.}
	\label{pure}
\end{figure*}
 \subsection{Overhead \& Complexity}
 Fig. \ref{overhead} illustrates the overhead and complexity of the network based on the different policies (i.e., centralized, distributed, and smart). We assume the set $ \{16, \: 4, \: 4\} $ as the number of information bits to transmit channel status, subcarrier indicators, and the transmission power in the feedback process. As we can see, the network overhead and complexity for the centralized structure is too high, which leads to numerous difficulties in the network, such as the high delay and low reliability. The overhead and complexity in the distributed network are lower, which makes the distributed architecture preferable for real-time reliable networks. Fig. 4 also shows that the complexity and overhead of the centralized structure are greater in the ultra-dense networks, which makes the centralized policy inefficient for a dense network. This is due to the fact that in the centralized scheme, the complexity and overhead grow linearly due to resource management being performed over all the transmission nodes of the network with information being exchanged between all nodes. In contrast, the complexity and overhead of the distributed scheme are inferior.~This indicates that the complexity gap between the two schemes is much more sensible in dense networks with many users. On the other hand, the overhead and complexity of the smart framework closely follow those of the centralized and distributed structures in low and ultra-dense networks, respectively. We remark that although for a single instance the smart decision selects one of the centralized/distributed approaches, as a result of the existence of different channels that are variable over the time, the overhead and complexity of the smart network on average would occur between two different strict centralized or distributed architecture.
\begin{figure*}[ht!]
	\centering
	\centerline{\includegraphics[width=0.7\linewidth]{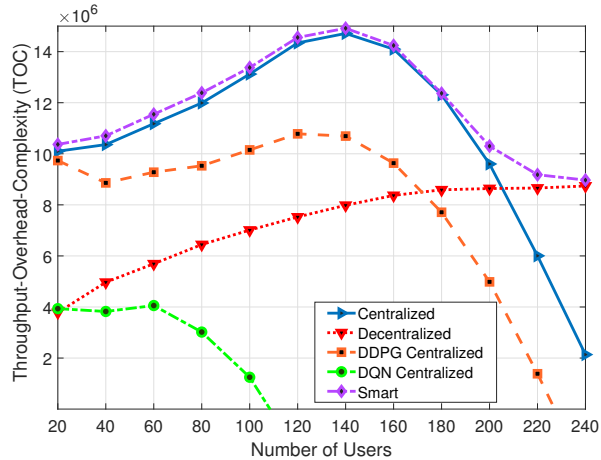}}
	\caption{TOC relative to the number of users.}
	\label{main_result}
\end{figure*}
\subsection{Original Average Data Rate}
In Fig. \ref{pure}, we compare the achievable data rate for the different proposed architectures. As we can see, the achievable data rate of the centralized scheme outperforms the others, because in this scheme the network's knowledge is collected at the centralized unit, which raises the network awareness and efficiency. The smart solution achieves a data rate intermediate between the centralized and distributed schemes. In particular, by increasing the network's traffic, the smart scheme at first follows the centralized structure and then approaches the distributed case. Further, as Fig. \ref{pure} shows, the achievable data rate for a huge number of users in all schemes is saturated due to the limited resources. We remark that since there are different channel gains which are variable over time, the smart algorithm achieves the data rate between the centralized and distributed results on average.
We can also observe in Fig. 5 that the efficiency of the soft actor-critic based algorithm is greater than that of the method. This makes sense since the soft actor-critic can be employed for the continuous variables while the DQN needs some discretized variables, that results in some performance loss. 
\subsection{New Performance Metric: Throughput Overhead Complexity (TOC)}
In this section, we introduce a new metric for evaluating system performance as Throughput Overhead Complexity (TOC) which considers the effect of overhead and complexity in the network. Fig. \ref{main_result} presents a comparison of TOC values for the centralized, distributed, and proposed smart network frameworks. It is assumed that the number of users in the network varies depending on the level of traffic, from low to heavy. As we can see in Fig. 6, the centralized network shows a moderate performance increase in terms of TOC when the traffic status is low while it experiences a sharp decrease in a dense network. In the distributed network, there is a moderate increase of performance in terms of TOC by increasing the load of the network. At $ |\mathcal{K}|=160 $ the throughput of the distributed structure exceeds that of the centralized one, and thus the decision-making algorithm here plays a vital role in the network.
 Another interesting observation in this figure is that the performance gain in terms of the TOC through the learning-based approaches is more efficient than that of DDPG method.
 This means that although the achieved data rate through the DDPG is higher than that of the ML solutions, the complexity of the DDPG in the centralized manner is very high, which deteriorates the performance gain in terms of TOC. 
\section{Conclusion}
In this paper, we proposed a hierarchical network management framework that adopts the best resource allocation policy in relation to changes in network status. The proposed intelligent framework is density aware, which guarantees the network's performance on the basis of the DRL algorithm. We investigated three different scenarios (i.e., fixed centralized, fixed distributed, and dynamic) for different levels of network traffic.~Simulation results showed that the proposed algorithm not only performs better than conventional learning methods in terms of TOC, but that it also outperforms both fixed centralized and distributed resource allocation policies.
% \begin{figure*}
% 	\centering
% 	\includegraphics[width=11.00cm]{figsac}
% 	\caption{Actor-critic Algorithm in an Agent.}
% 	\label{DDRLF}
% \end{figure*}

%%===================================================================
%\input{Conclusion}

\bibliographystyle{IEEEtran}
\bibliography{Bibliography}
\end{document}